\documentclass[aps,prl,reprint,superscriptaddress,floatfix,amsmath,showpacs]{revtex4-1}

\usepackage{graphicx}
\usepackage{amsmath}
\usepackage{textcomp}
\usepackage{xcolor}
\usepackage{hyphenat}
\usepackage[normalem]{ulem} 
\usepackage[ps2pdf,colorlinks=true,allcolors=blue,breaklinks=true]{hyperref} 

\RequirePackage{caption}
\DeclareCaptionLabelSeparator{textbar}{.}
\begin{document}

\title{Unconventional spin currents in magnetic films}

\author{Dmytro\,A.\,Bozhko}
\email{bozhko@physik.uni-kl.de}
\affiliation{Fachbereich Physik and Landesforschungszentrum OPTIMAS, Technische Universit\"at Kaiserslautern, 67663 Kaiserslautern, Germany}
\affiliation{School of Engineering, University of Glasgow, Glasgow G12 8LT, United Kingdom}

\author{Halyna\,Yu.\,Musiienko-Shmarova}
\affiliation{Fachbereich Physik and Landesforschungszentrum OPTIMAS, Technische Universit\"at Kaiserslautern, 67663 Kaiserslautern, Germany}

\author{Vasyl\,S.\,Tiberkevich}
\affiliation{Department of Physics, Oakland University, Rochester, Michigan 48309, USA}

\author{Andrei\,N.\,Slavin}
\affiliation{Department of Physics, Oakland University, Rochester, Michigan 48309, USA}

\author{Ihor\,I.\,Syvorotka}
\affiliation{Department of Crystal Physics and Technology, SRC ``Carat'', 79031 Lviv, Ukraine}

\author{Burkard\,Hillebrands}
\email{hilleb@physik.uni-kl.de}
\affiliation{Fachbereich Physik and Landesforschungszentrum OPTIMAS, Technische Universit\"at Kaiserslautern, 67663 Kaiserslautern, Germany}

\author{Alexander\,A.\,Serga}
\affiliation{Fachbereich Physik and Landesforschungszentrum OPTIMAS, Technische Universit\"at Kaiserslautern, 67663 Kaiserslautern, Germany}

\begin{abstract}
\noindent\textbf{
\nohyphens{
A spin current --- a flow of spin angular momentum --- can be carried either by spin polarised free electrons or by magnons, the quanta of a moving collective oscillation of localised electron spins --- a spin wave.
Traditionally, it was assumed, that a spin wave in a magnetic film with spin-sink-free surfaces can transfer energy and angular momentum only along its propagation direction. In this work, using Brillouin light scattering spectroscopy in combination with a theory of dipole-exchange spin-wave spectra, we show that in obliquely magnetized free magnetic films the in-plane propagation of spin waves is accompanied by a transverse spin current along the film normal without any corresponding transverse transport of energy.
}}
\end{abstract}

\maketitle

\everypar{\looseness=-1}

Spin waves (or their quanta -- magnons) can be used as carriers of a spin current in spintronics \cite{Chumak2015,Yu2018} and magnonics \cite{Serga2010,Kruglyak2010,Khitun2010} signal processing devices, as they enable transport of energy and spin angular momentum over long distances along their propagation direction \cite{Kajiwara2010,Liu2018,Chen2017}. This has been used for concepts in view of novel spintronic data processing applications, such as planar spin-wave conduits \cite{Wang2018,Wagner2016}, magnonic crystals \cite{Chumak2017,Lisenkov2015}, logic elements, such as spin-wave majority gates \cite{Fischer2017}, and nonlinear spin-wave devices, such as magnon transistors \cite{Chumak2014}. Many of these novel elements operate in the dipole-exchange region of a spin-wave spectrum, where the strength of magnetic dipole-dipole and local electrostatic exchange interactions between electron spins are comparable. Most experimental studies of dipole-exchange spin waves have been performed using in-plane magnetized films, where the spin-wave propagation in the long-wavelength part of the spectrum is highly anisotropic due to the influence of the dipolar interaction, and exotic phenomena such as spin-wave caustics \cite{Schneider2010} may appear. Furthermore, the interplay between dipole and exchange interactions in this magnetization geometry creates prerequisites for fundamental phenomena such as magnon Bose-Einstein condensation \cite{Demokritov2006,Serga2014}, magnonic supercurrents \cite{Bozhko2016,Nakata2014,Nakata2015}, formation of hybrid magnetoelastic bosons \cite{Bozhko2017}, spin superfluidity \cite{Tserkovnyak2014, Sun2016,Skarsvag2015}, and more. In spin-sink-free in-plane magnetized magnetic films, the energy and spin angular momentum are transported by spin-wave modes along the film plane, and the transverse (along the film normal) profiles of the spin-wave modes are standing waves, which do not exhibit a net transport of energy or spin angular momentum. 
However, if a spin sink, i.e. an overlayer with strong spin-orbit coupling (such as Pt or Ta) is attached to the film, the absorption of magnons by such a spin sink leads to the appearance of spin and energy flows along the film normal \cite{Chumak2012,Kajiwara2011}. It allows for the detection of propagating spin waves in a large range of wavelengths \cite{Chumak2012} and, thus, plays an important role in the development of magnon spintronics towards utilization of short-wavelength dipolar-exchange and exchange spin waves in nano-sized devices \cite{Chumak2015}.

\begin{figure}[b!]
	\includegraphics[width=0.9\columnwidth]{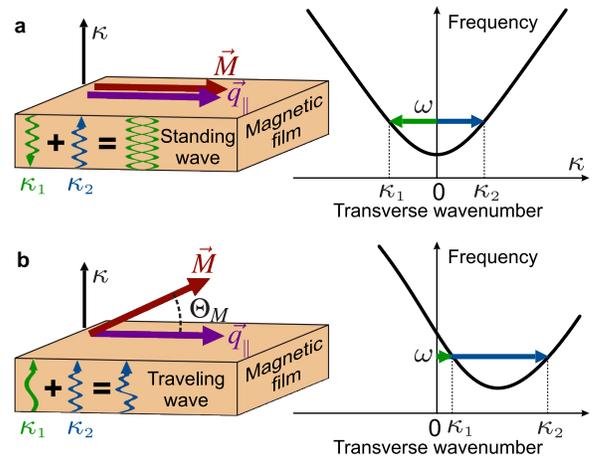}
	\caption{\label{FigExplanation} \textbf{Standing- and travelling mode profiles.} Dependence of the spin-wave frequency on the transverse (to the direction of wave propagation and along the film normal) wavenumber $\kappa$ in a ferromagnetic film for different angles $\Theta_M$ between the static magnetization and the direction of the wave propagation $\textbf{\textit{q}}_{_\parallel}$: \newline \textbf{a}, $\Theta_M = 0^\circ$; \textbf{b}, $\Theta_M > 0^\circ$.}
\end{figure}

Here we report, that, in contrast with the case of in-plane magnetization, in obliquely magnetized films, the transverse profiles of in-plane-propagating dipole-exchange spin waves are only quasi-standing, and may allow for the transport of a spin angular momentum along the film normal even in the absence of any spin sink. Noteworthy, in the case of no spin absorption at the film boundaries, such currents abstain from a net transport of energy. These unconventional spin currents are crucially important for applications in modern spintronics, where they can substantially influence the effects of spin pumping and spin transfer torque at a film interface.

First we consider a gedanken experiment, such as schematically shown in Fig.\,\ref{FigExplanation}b, where a spin wave travels in-plane in $+\textbf{\textit{q}}_{_\parallel}$ direction. Such a scenario can be realized by using an antenna structure to generate the spin waves, positioned left to the spot, where we detect the spin waves in a Brillouin light scattering experiment, see also discussion below. Similarly to the case of conventional electromagnetic or optical waveguides, the profile along the film normal of a spin-wave mode propagating in a film plane with a wavevector $\textbf{\textit{q}}_{_\parallel}$ can be described as a superposition of two partial plane waves with transverse wavenumbers $\kappa_{\mathrm{1}}$ and $\kappa_{\mathrm{2}}$ at the same frequency $\omega = \omega(\textbf{\textit{q}}_{_\parallel}, \kappa_{\mathrm{1}}) = \omega(\textbf{\textit{q}}_{_\parallel}, \kappa_{\mathrm{2}})$ (see Fig.\,\ref{FigExplanation}). The in-plane spin-wave dispersion, however, is anisotropic with a minimum frequency corresponding to the wave propagation along the static magnetization (see Methods). For the case of an in-plane magnetization, this minimum is achieved at $\kappa = 0$ and the dependence of the frequency on the wavenumber $\kappa$ is symmetric: two partial waves are contra-propagating with equal-by-magnitude but opposite-by-sign wavenumbers $\kappa_{\mathrm{1}} = - \kappa_{\mathrm{2}}$ (see Fig.\,\ref{FigExplanation}a). Therefore, the corresponding mode profile is represented by a classical standing wave pattern (see sketch in Fig.\,\ref{FigExplanation}a). The situation changes when the sample is magnetized obliquely (see Fig.\,\ref{FigExplanation}b). In this case, the minimum spin-wave frequency corresponds to a certain $\kappa \ne 0$ value and the dependence $\omega = \omega(\textbf{\textit{q}}_{_\parallel}, \kappa$) is asymmetric. Thus, at some frequencies the mode profiles are formed by co-propagating partial waves. Their superposition results in a travelling wave pattern, which carries angular momentum and, thus, can be treated as a transversal spin current $i_\mathrm{s}$ directed along the film normal.

It is worth noting that in the framework of the model of a free film (in the case of absence of spin sinks on the film surfaces) there is no energy flow along the film normal. Moreover, regarding the in-plane wavevector $+\textbf{\textit{q}}_{_\parallel}$, which is assumed here to be set by the excitation scheme, spin-wave modes with positive and negative wavevectors $\pm\textbf{\textit{q}}_{_\parallel}$ carry opposite transversal spin currents. Thus, in the case of a symmetric spin wave excitation or under thermal equilibrium, the net current $i_\mathrm{s}$ is zero. In view of these facts the question arises if it would be possible, in principle, to reveal the existence of such a current in a magnetic film, which is free from any type of a spin sink? 

Here, we show that the investigation of this fundamental phenomenon is possible in a Brillouin light scattering (BLS) experiment. The idea behind is that any modification of the spin-wave profiles along the film normal changes the BLS cross-section regardless the positive or negative orientation of the wavevector $\textbf{\textit{q}}_{_\parallel}$. In the present work, we use this fact to demonstrate the predicted emergence of the unconventional transversal spin current in an obliquely magnetized thermally excited YIG sample.

\begin{figure}[t!]
	\includegraphics[width=0.8\columnwidth]{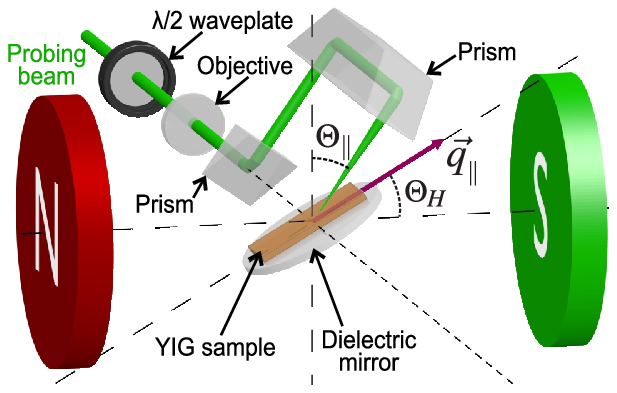}
	\caption{\label{FigSetup} \textbf{Experimental BLS set-up with wavevector resolution.} A probing laser beam is focused onto the sample, which is placed on a dielectric mirror. A system of prisms allows for variation of the light incidence angle without shift in the focal spot position. The inelastically scattered light is collected in the direction of the incident beam, thus realizing wavevector resolution (see Methods). The dielectric mirror is mounted on a piezo-driven rotary stage, which allows for the controlled variation of the angle of magnetization $\Theta_H$ in the range from $0^\circ$ (in-plane magnetization) to $90^\circ$ (out-of-plane magnetization). }
\end{figure}

\begin{figure*}[t!]
	\includegraphics[width=0.8\textwidth ]{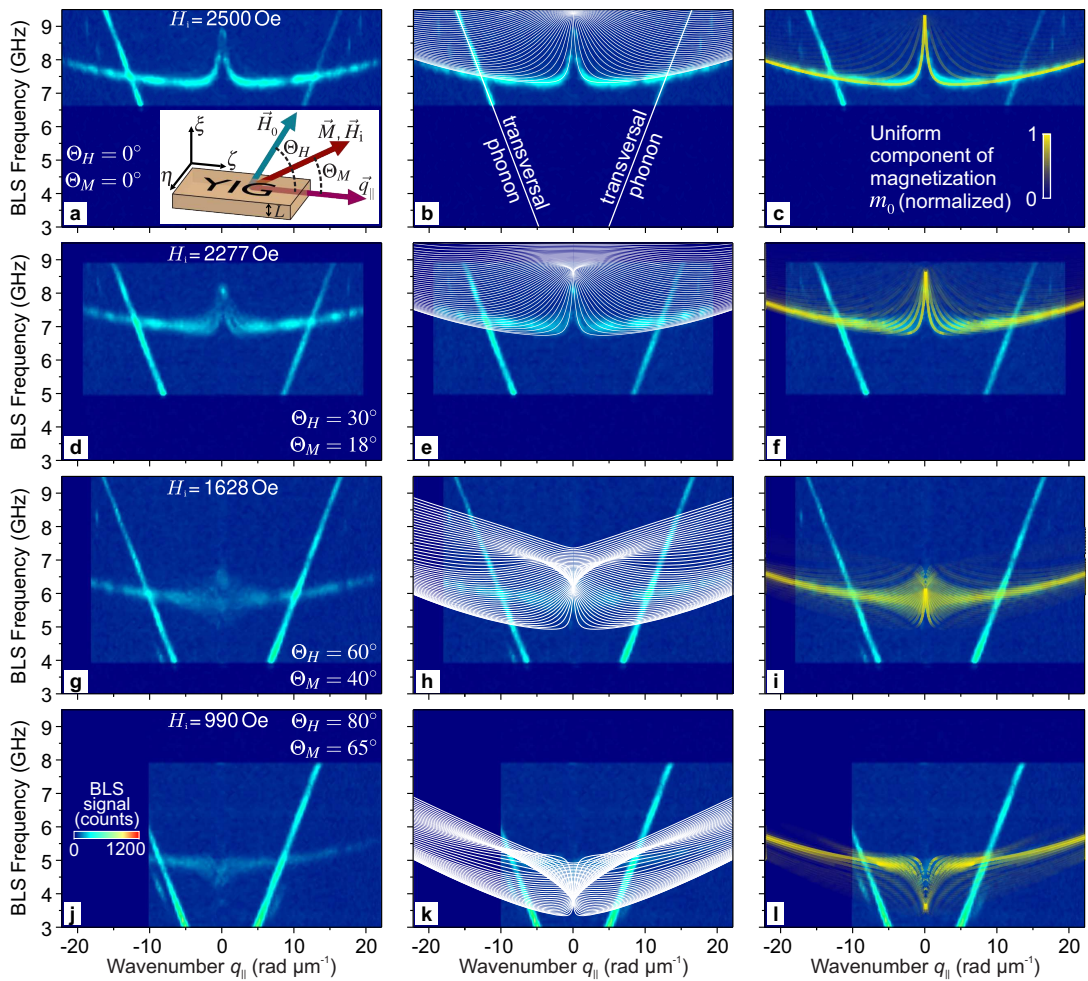}
	\caption{\label{FigExperiment} \textbf{Spectra.} Thermal spin-wave spectra in a $5.6\,\mu\text{m}$ - thick YIG film sample measured using a BLS set-up with wavevector resolution (see Fig.\,\ref{FigSetup}) at different magnetization angles $\Theta_H$. Shown is the measured BLS intensity (colour code) as function of frequency (vertical scale) and transferred wavevector (horizontal scale). Only the Stokes component of the spectrum of inelastically scattered light is shown. The inset in \textbf{a} shows the magnetization geometry. \textbf{a-c}, $\Theta_H = 0^\circ$; \textbf{d-f}, $\Theta_H = 30^\circ$; \textbf{g-i}, $\Theta_H = 60^\circ$; \textbf{j-l}, $\Theta_H = 80^\circ$. The panels in the left column demonstrate the raw measured spin-wave spectra. In the panels presented in the middle column the same data are shown together with the theoretically calculated dispersion curves for the first 50 thickness modes. The images in the right column show the experimentally measured spectra together with the calculated curves of the BLS spectral visibility defined as the normalized spatially uniform component of the profile along the film normal of a particular spin-wave mode.
	}
\end{figure*}

The spin-wave spectra are studied at room temperature in a low-damping ferrimagnetic film of Yttrium Iron Garnet (YIG, $\mathrm{Y_{3}Fe_{5}O_{12}}$)\cite{Cherepanov1993, Dubs2017} by means of BLS spectroscopy with frequency and wavevector resolution (see Fig.\,\ref{FigSetup}). The YIG film sample of thickness of $L=5.6\,\mu\text{m}$ is biased by the magnetic field $H_0=2500\,\text{Oe}$, which is sufficient to magnetize the sample with the saturation magnetization of $4\pi M_\mathrm{s} = 1750\,\text{G}$ even when the bias field is perpendicular to the YIG film surface.

The experimental set-up, which consists of a YIG film and a wavevector-resolving BLS system, is schematically shown in Fig.\,\ref{FigSetup} (for a detailed description of the set-up and techniques see Methods). The sample is placed on top of a dielectric mirror, which ensures a perfect reflection of the scattered light, and, thus, significantly enhances the wavenumber selectivity and sensitivity of the experimental set-up allowing detection of thermal spin waves. In addition, this mirror, being dielectric, does not change the magnetic boundary conditions at the YIG film surface and thus, does not modify the spectra of the spin waves as it can occur in the conventional case of a  metal mirror \cite{Wames1969, Mruczkiewicz2013}. The developed set-up is placed inside a magnet gap and allows for spin-wave measurements in the full range of the magnetization angle $\Theta_H$ from $0^\circ$ (in-plane magnetization) to $90^\circ$ (out-of-plane magnetization).

Next, we study the thermally excited spin-wave spectrum as a function of $\Theta_H$. With increasing $\Theta_H$ both the magnitude and direction of the internal bias magnetic field $\textbf{\textit{H}}_\mathrm{i}$ change, because of the influence of the demagnetizing field. The angle $\Theta_M$ of the internal magnetization $\textbf{\textit{M}}$ and the magnitude of the internal magnetic field $H_{\mathrm{i}}$ can be found from the electrodynamic boundary conditions at the surface of the YIG film \cite{Gurevich1996}. The values of $H_{\mathrm{i}}$ and $\Theta_M$ calculated for the particular values of $\Theta_H$ used in our experiments are given in Fig.\,\ref{FigExperiment}a, d, g, and j.

The observed thermal spectra of dipole-exchange spin waves superimposed by phonon branches are shown in Fig.\,\ref{FigExperiment}. Panels a, d, g, and j in the left column of Fig.\,\ref{FigExperiment} show maps of the raw BLS intensity obtained from the room-temperature populations of magnons and phonons propagating in the YIG film along the projection of the bias magnetic field onto the film plane as functions of the magnon (or phonon) wavevector $\textbf{\textit{q}}_{_\parallel} \parallel \textbf{\textit{H}}_0$. \looseness=0

For the in-plane magnetization case ($\Theta_H = 0^\circ$, Fig.\,\ref{FigExperiment}a), the experimentally visible spin-wave spectrum consists mainly of the lowest mode of the dipole-exchange back\-ward volume waves \cite{Stancil2009}. The interplay of dipole and exchange interactions is clearly visible: the initial dipolar decrease of the spin-wave frequency in the long-wavelength section of the spectrum is superseded by the exchange-determined frequency increase at larger wavenumbers $\pm q_{_\parallel}$. The steep straight dispersion branches also seen in the data correspond to the transverse acoustic phonons \cite{Bozhko2017,Rueckriegel2014,Flebus2017,Cornelissen2017,Kikkawa2016}.

The variation of the magnetization angle $\Theta_H$ leads to a modification of the observed spectra (see panels d, g, and j in Fig.\,\ref{FigExperiment}). For $\Theta_H = 30^\circ$ shown in Fig.\,\ref{FigExperiment}d one can see that several dispersion branches with negative group velocities $v_\mathrm{g}<0$ become visible in the dipolar (long wavelength) region of the spectrum ($|q_{_\parallel}|<1\,\text{rad}\,\mu\text{m}^{-1}$). With further increase in $\Theta_H$ the contrast of the BLS map in this region decreases, while the exchange magnons and transverse phonons are still clearly visible. The shape of the blurred part of the spin-wave spectrum (see Fig.\,\ref{FigExperiment}g near $q_{_\parallel}=0$) gives a hint to the presence of modes with both positive and negative group velocities $v_\mathrm{g}$ in the spin-wave spectrum (both backward and forward types of dipole-exchange magnons \cite{Stancil2009,Slavin1988}). At the angle of magnetization of $\Theta_H = 80^\circ$ only magnon modes with $v_\mathrm{g}\gtrsim 0$ are seen with rather low contrast in the dipolar area of the spectrum, and all these modes merge into a single exchange-dominated mode at large wavenumbers ($|q_{_\parallel}|>1\,\text{rad}\,\mu\text{m}^{-1}$). No spin waves are detected at $\Theta_H = 90^\circ$: the intensity of the spin-wave maps becomes smaller with increase of the magnetization angle, and approaches zero when the magnetization becomes exactly perpendicular to the film plane due to the corresponding decrease in the BLS cross-section \cite{Cardona2000}.

Since our BLS set-up is sensitive only to the modes with rather uniform profiles along the film normal, the observed blurring of the spin-wave spectrum for $\Theta_H \ne 0^\circ$ suggests a substantial transformation of the spin-wave mode profiles when compared to the case of in-plane magnetization. To understand this transformation we further developed the formalism described in Ref.\,\cite{Kalinikos1986} (see Methods). Both the dispersion characteristics $\omega(q_{_\parallel})$ and the profiles along the film normal $\mathbf{m}(\xi)$ of the spin-wave modes were calculated and are determined as a function of the in-plane wavenumber $\pm q_{_\parallel}$ (for the used coordinate system please see inset in Fig.\,\ref{FigExperiment}a).  

The theoretically calculated spin-wave dispersion characteristics plotted by white lines are presented together with the experimental data in the central column of Fig.\,\ref{FigExperiment} (panels b, e, h, k). It can be clearly seen that the calculation describes the fundamental spin-wave mode in the case of in-plane magnetization very well (Fig.\,\ref{FigExperiment}b), but the higher-order modes present in the theory are not visible in the experiment. At magnetization angles $\Theta_H > 0^\circ$ the situation becomes worse, since many of the theoretically predicted spin-wave modes are not observed in our BLS experiment (see panels e, h, and k in Fig.\,\ref{FigExperiment}). This apparent discrepancy is related to the aforementioned fact that the BLS cross-section is sensitive only to the modes having a uniform component in their profiles along the film normal. Thus, the spin-wave modes with profiles oscillating about zero along the film normal do not contribute to the BLS signal.

\begin{figure}[t!]
	\includegraphics[width=0.8\columnwidth]{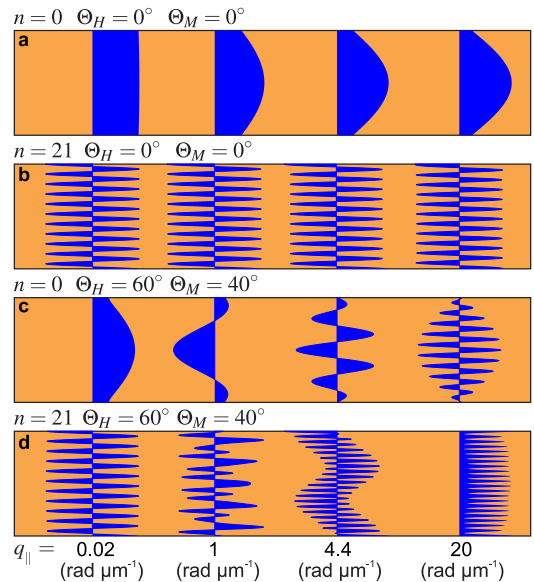}
	\caption{\label{FigProfiles} \textbf{Mode profiles.} \textbf{a-d} Numerically calculated profiles along the film normal of the spin-wave modes in a $5.6\,\mu\text{m}$ thick YIG film sample.}
\end{figure}

The calculated profiles of two spin-wave modes are exemplarily shown in Fig.\,\ref{FigProfiles} for a few selected wavenumbers $q_{_\parallel}$. The spatially uniform zero Fourier component of each of such a profile is directly proportional to the sensitivity of the BLS set-up to the corresponding spin-wave mode. The magnitudes of this component are presented in the right column of Fig.\,\ref{FigExperiment} (panels c, f, i, l) by yellow lines jointly with the BLS intensity maps. As can be seen, these yellow lines agree very well with the experimental data (compare panels c, f, i, and l in Fig.\,\ref{FigExperiment} to panels a, d, g, and j in the same figure). The fundamental ($n=0$) mode has a perfectly uniform profile along the film normal at small $(q_{_\parallel}L <1)$ in-plane wavenumbers $q_{_\parallel}=0.02\,\text{rad}\,\mu\text{m}^{-1}$, and is therefore very visible in the BLS experiment. With the increase in $q_{_\parallel}$ the effective pinning in the profile along the film normal of this mode gradually increases. This pinning is not at all related to any surface anisotropy at the film boundaries, but has a purely dipolar character \cite{Guslienko2002, Guslienko2005}. Note, also that the higher order (e.g. $n=21$) modes have oscillating profiles along the film normal, and therefore, cannot be detected in the BLS experiment. 

The mode profiles in an obliquely magnetized magnetic film (for illustration, we consider here the case of $\Theta_H =60^\circ$) have rather unusual shapes (see Fig.\,\ref{FigProfiles}c and \ref{FigProfiles}d). The profile along the film normal of the $n=0$ mode at relatively small wavenumber values $q_{_\parallel}=0.02\,\text{rad}\,\mu\text{m}^{-1}$ is rather uniform in comparison to the profiles shown in Fig.\ref{FigProfiles}b, and this mode is therefore clearly visible in the BLS experiment (similar to the situation in Fig.\ref{FigProfiles}a). However, with increase in $q_{_\parallel}$, the mode profile acquires more and more oscillations along the film normal, leading to a lower visibility of these modes in the experiment (see Fig.\,\ref{FigExperiment}g). At $q_{_\parallel}>1\,\text{rad}\,\mu\text{m}^{-1}$ the visibility of these modes vanishes completely. The opposite situation occurs for the higher-order mode with $n=21$ (see Fig.\,\ref{FigProfiles}d). The profile along the film normal of such a mode, that is purely oscillating at $q_{_\parallel}=0.02\,\text{rad}\,\mu\text{m}^{-1}$, continues to oscillate at higher values of the in-plane wavenumber but, surprisingly enough, also possesses a non-zero uniform component (see Fig.\,\ref{FigProfiles}d for $q_{_\parallel}=20\,\text{rad}\,\mu\text{m}^{-1}$). It is worth to note, that the above described transformation of the mode profiles along the film normal is not exclusively related to the case of a plain YIG film, but can also explain the effects observed in Permalloy micro-structured nanodots \cite{Verba2016,Kakazei2012} and magnetic multilayers \cite{Stamps1991}. \looseness=0

The unusual profiles of the spin-wave modes presented in Figs.\,\ref{FigProfiles}c and \ref{FigProfiles}d find a natural qualitative explanation in the framework of the proposed model of transversal spin currents.

In the case of an in-plane magnetized film (Fig.\,\ref{FigExplanation}a), which is characterized by the symmetric dependence $\omega(\kappa)$ for positive and negative $\kappa$ directions, the mode profiles demonstrate perfect standing wave patterns along the normal of the film (see Fig.\,\ref{FigProfiles}a,b). The fundamental mode with the lowest possible frequency has $|\kappa_{\mathrm{1,2}}| \approx 0$, i.e., its profile is quasi-uniform (see Fig.\,\ref{FigProfiles}a). With increasing frequency, the transversal wavenumbers $|\kappa_{\mathrm{1,2}}|$ also increase, and therefore increases the number of nodes in the mode's profile along the film normal (the mode's index $n$) as it is visible in Fig.\,\ref{FigProfiles}b.

The situation is different for an obliquely magnetized film. Due to the anisotropic character of the dipolar magnetic interaction \cite{Wames1969,Kalinikos1986} the dependence $\omega(\kappa)$ is no longer symmetric (see Fig.\,\ref{FigExplanation}b).  Now, for spin waves with relatively long wavelength, the minimum frequency is achieved when their total wavevector $\textbf{\textit{q}}$ is parallel to the internal magnetic field, i.e., for $\kappa = \kappa_{\mathrm{min}} \approx q_{_\parallel} \tan\Theta_M$. Therefore, in this case, the fundamental mode will be formed by two partial waves $\kappa_{\mathrm{1}} \approx \kappa_{\mathrm{2}} \approx \kappa_{\mathrm{min}}$ and possesses approximately $\kappa_{\mathrm{min}}L/\pi \approx (q_{_\parallel}L/\pi) \tan\Theta_M$ nodes along the film normal. We would like to underline the fact that in the shown case both partial waves have transverse wavevectors of same direction, so the resultant mode has a quasi-travelling character, but the group velocities of the partial waves have opposite signs, which means that the mode does not transfer energy along the film normal. The increase of the number of nodes of the fundamental mode with an increase in $q_{_\parallel}$ is clearly seen in Fig.\,\ref{FigProfiles}c. With the increase in the mode number $n$ the transversal wavenumber $\kappa_{\mathrm{1}}$ decreases, while $\kappa_{\mathrm{2}}$ increases (see Fig.\,\ref{FigExplanation}b), producing the non-trivial patterns seen in Fig.\,\ref{FigProfiles}d. In particular, for a certain higher-order mode (or, in the case of a fixed mode number $n$, for a particular in-plane wavevector $q_{_\parallel}$) $\kappa_{\mathrm{1}}\approx 0$, and the mode attains a comb-like profile seen in Fig.\,\ref{FigProfiles}d for $q_{_\parallel} = 20\,\text{rad}\,\mu\text{m}^{-1}$.

The above considerations lead to the essential conclusion which is that the profiles along the film normal of the magnon modes propagating in an obliquely magnetized magnetic film do not fully form standing wave patterns, but have a travelling character. This conclusion is further confirmed by a direct visualization of the time evolution of the mode profiles shown in the Supplementary Movie 1.

This unconventional travelling nature of the magnon mode's profiles has important implications in the field of modern spintronics. Particularly, by direct evaluation, one can show that these modes carry a non-zero intrinsic exchange spin current
\begin{equation}\label{spincurrent}
i_\mathrm{s} \propto |\textbf{\textit{m}} \times \partial\textbf{\textit{m}}^*/\partial\xi|
\end{equation}
in the direction perpendicular to the film plane. From the point of view of the symmetry considerations, the appearance of the perpendicular spin current is associated with the symmetry breaking induced by the oblique magnetization and the direction of the in-plane wavevector of the spin waves, $\mathbf{q}_{_\parallel}$. Reversing either the direction of the bias magnetic field or the direction of $\mathbf{q}_{_\parallel}$ will reverse the sign of the transverse spin current. In particular, the spin waves with $q_{_\parallel} \simeq 0$ always carry zero perpendicular current, so the effects related to this current require travelling spin waves, and cannot be seen in a ferromagnetic resonance kind of experiments. 

\begin{figure}[t!]
	\includegraphics[width=0.9\columnwidth]{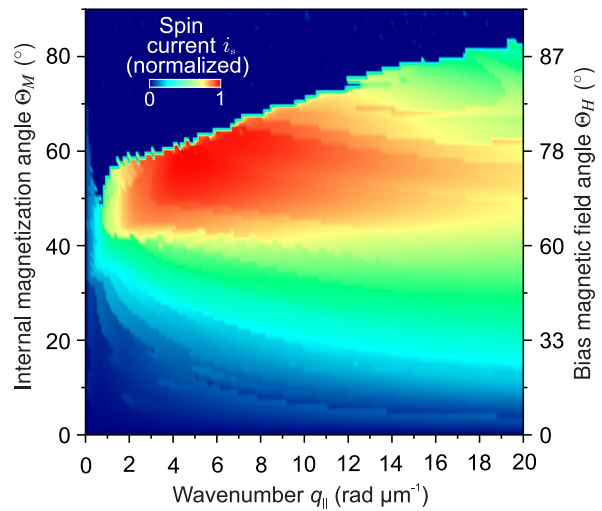}
	\caption{\label{FigSpinCurrent} \textbf{Spin current.} Colour coded intensity of spin current $i_\mathrm{s}$ calculated using Eq.\,(\ref{spincurrent}) as a function of the in-plane wavenumber $q_{_\parallel}$ and magnetization angle $\Theta_M$ for the modes with quasi-uniform distribution along the film normal.}
\end{figure}

It should be remarked, that in the state of thermal equilibrium, the magnon modes with wavevectors $+q_{_\parallel}$ and $-q_{_\parallel}$ are equally populated, and the transverse spin currents they carry mutually compensate each other. Therefore, the intrinsic transverse spin current does not have any direct influence on the thermal equilibrium properties of a magnetic film. 

If a spin wave is excited in a certain direction, lets assume in $+\zeta$ direction, i.e., we only have $+q_{_\parallel}$, the effect of a transverse spin current will manifest itself explicitly. In this scenario we may find different spin pumping efficiencies at the opposite surfaces of the film. This effect, in particular, can be used for the development of non-reciprocal spin current transmission lines based on propagating spin waves. The contribution of the spin current $i_\mathrm{s}$ for the modes with quasi-uniform profile along the film normal is of particular interest (e.g. all profiles in Fig.\,\ref{FigProfiles}a and the profile at $q_{_\parallel}=20\,\text{rad}\,\mu\text{m}^{-1}$ in Fig.\,\ref{FigProfiles}d). These modes can be excited, for example by micro-structured microwave antennas. We have examined the $\Theta_H$ versus $q_{_\parallel}$ parameter space regarding the transversal spin current efficiency. The results of numerical calculations of the spin current magnitude using Eq.\,(\ref{spincurrent}) are shown in Fig.\,\ref{FigSpinCurrent}. As can be seen, the maximal transversal spin current is achieved at approximately $\Theta_M \simeq 55^\circ$ and $q_{_\parallel} \simeq 5\,\text{rad}\,\mu\text{m}^{-1}$, which corresponds to the wavelength of $1.2\,\mu\text{m}$ and fits well to the requirements of modern nano-scale magnonic devices. As the described spin currents along the film normal originate from the interplay between dipolar and exchange interactions, they vanish in the case of pure exchange waves as well as for pure dipolar waves. As expected, $i_\mathrm{s}$ weakens with an increase in the in-plane wavenumber $q_{_\parallel}$ and becomes zero at the limiting values ($0$ and $90^\circ$) of the magnetization angle $\Theta_M$ (see Fig.\,\ref{FigSpinCurrent}).

The presence of these unconventional spin currents may be directly detected in the following proposed experiment. In the case of oblique magnetization, spin waves propagating in opposite in-plane directions should have different spin pumping efficiencies at the surfaces of the magnetic film covered with a few nanometer-thick non-magnetic heavy metal. The difference in the values of spin currents injected by the contra-propagating waves to such a spin sink can be measured as a corresponding difference in the inverse spin Hall voltages generated across the metal layer. However, the excitation of short-wavelength spin waves ($|q_{_\parallel}|>4\,\text{rad}\,\mu\text{m}^{-1}$), for which this effect is expected to be sufficiently high (see Fig.\,\ref{FigSpinCurrent}), is rather difficult task due to the absence of reliable unidirectional spin-wave emitters. Development of such emitters is a work in progress \cite{Wintz2016,Davies2016,Dieterle2019}.

We believe that these interesting and non-trivial properties of the spin-wave modes in obliquely magnetized magnetic films will be used in the future to control the effects of spin pumping and to generate spin currents in nano-scale spintronic signal processing and signal-generating devices.
Moreover, similar effects should also be observed in photonic and acoustic waveguide structures made of anisotropic materials, in the case when the directions of the main crystallographic axis does not coincide with the direction of wave propagation.

\setlength{\smallskipamount}{1.0mm}

\bigskip
{\fontfamily{phv}
	\fontsize{11pt}{10.0pt}
	\selectfont
	\noindent \textbf{Acknowledgements}} 

\noindent Support by the Deutsche Forschungsgemeinschaft within the Research Units TRR 49 (Project A07 INST 161/544-3)  and TRR 173 ``Spin+X'' (Projects B01 INST 248/205-1) as well as financial support by the European Research Council within the Advanced Grant ``Supercurrents of Magnon Condensates for Advanced Magnonics'' is gratefully acknowledged. This work was also supported in part by the grants Nos. EFMA-1641989 and ECCS-1708982 from the National Science Foundation of the USA, and by the DARPA M3IC grant under the contract W911-17-C-0031. We are also indebted to R.V. Verba and P. Pirro for discussions. \looseness=-1

\newpage

\bigskip

{\fontfamily{phv}
	\fontsize{11pt}{10.0pt}
	\selectfont
	\noindent\textbf{Methods}}

\smallskip
\noindent\textbf{Sample.}
The yttrium iron garnet (YIG, $\mathrm{Y_{3}Fe_{5}O_{12}}$) sample is 5\,mm long and 1\,mm wide. The single-crystal YIG film of 5.6\,$\mu$m thickness has been grown in the (111) crystallographic plane on a gadolinium gallium garnet (GGG, $\mathrm{Gd_{3}Ga_{5}O_{12}}$) substrate by liquid-phase epitaxy.

\smallskip
\noindent\textbf{Brillouin light scattering spectroscopy.} The spin-wave spectrum was visualized by means of wavevector-resolved BLS spectroscopy\cite{Sandweg2010} in back-scattering geometry. 
\noindent Brillouin light scattering spectroscopy can be understood as the diffraction of the probing light from a moving Bragg grating produced by a spin-wave mode \cite{Demokritov2001}. As a result, some portion of the scattered light is shifted in frequency by the frequency of this mode. In addition, the diffraction from the grating leads to a transfer of momentum during this process. For example, the in-plane component of the wavevector $\textbf{\textit{q}}_\mathrm{L}$ of the incident light is inverted by a spin-wave mode propagating along the projection of a probing beam on a thin magnetic film if the spin-wave wavenumber $q$ satisfies the Bragg condition $q = -2q_\mathrm{L} \sin(\Theta_{\parallel})$, where $\Theta_{\parallel}$ is the angle of laser light incidence. By changing the angle $\Theta_{\parallel}$, wavevector selection of in-plane spin waves with wavevector $\textbf{\textit{q}}$ can be implemented. 

\noindent In our experiment, the probing laser beam of $532\,\text{nm}$ wavelength generated by a single-mode solid-state laser is focused by a precise objective onto the YIG film sample. The probing beam is steered by a system of prisms, enabling rotation of the probing laser beam around the sample. The opto-mechanical system is precisely aligned to keep the position of the probing spot at the same place at the sample surface while changing the angle of incidence $\Theta_{\parallel}$. In such a set-up, the rotation angle $\Theta_{\parallel}$ (see Fig.\,\ref{FigSetup}) and, consequently, the magnitude of the probed magnon wavevector $\textbf{\textit{q}}$ is limited only by the wavelength $\lambda$ of the probing laser light (for $\lambda=532\,\text{nm}$, $q^{\mathrm{max}}_{_\parallel}=2 q_{\mathrm{L}} = 23.6\,\text{rad}\,\mu\text{m}^{-1}$), and is not constrained by geometry of the magnet, as it was the case in Ref.\,\cite{Sandweg2010}. The polarization of the incident light was held parallel to the surface of the sample by utilizing a $\lambda/2$ wave plate, which is rotated simultaneously with the rotation of the prisms. The whole beam steering system is placed directly between the poles of the electromagnet, ensuring high field uniformity and stability. The sample is placed with the YIG side on top of a broadband dielectric mirror (commercially available ``Thorlabs'' E02 multilayer mirror for the 400-700\,nm light wavelength). As it was checked in test experiments, this mirror does not affect the phonon and spin-wave dispersions. Its role is to reflect the inelastically scattered light, whose in-plane wavevector component is inverted by a spin-wave mode, back through the objective for further frequency analysis. The dielectric mirror is fixed on a piezo-driven non-magnetic rotary stage, which was used to change the angle $\Theta_H$ between the bias magnetic field and the film surface.

\noindent The light collected by the objective is directed to a multipass tandem Fabry-P\'{e}rot interferometer \cite{Sandercock1981,Mock1987,Hillebrands1999} for frequency selection. At the output of the interferometer a single photon counting avalanche diode detector is placed. The output of the detector is connected to a counter module. Every time the detector registers a photon, this event is recorded to a database which collects the number of arrived photons. The frequency of the interferometer's transmission is also recorded, thus providing frequency information for each detected photon. \looseness=-1

\smallskip
\noindent\textbf{Simplified model of dipole-exchange spin-wave spectrum.} Considering the Landau-Lifshitz and magnetostatic Maxwell equations inside a magnetic film one can show that a propagating magnon mode can be represented by the sum of several partial plane waves with a total wavevector $\textbf{\textit{q}} = q_{_\parallel} \textbf{\textit{e}}_\zeta + \kappa \textbf{\textit{e}}_\xi$, where $\textbf{\textit{e}}_\zeta$ and $\textbf{\textit{e}}_\xi$ are unit vectors along the in-plane propagation direction and perpendicular to the film plane, respectively (see Fig.\,\ref{FigExplanation} and inset in Fig.\,\ref{FigExperiment}a). The relation between the total wavevector $\textbf{\textit{q}}$ and the mode frequency $\omega$ is the same as in the case of an infinite magnetic medium:
\begin{equation}\label{dispersion}
\omega^2 = (\omega_H + \omega_M \lambda_{\mathrm{ex}}^2 q^2)
(\omega_H + \omega_M \lambda_{\mathrm{ex}}^2 q^2 + \omega_M \sin^2\theta_q) \,.
\end{equation}
Here $\omega_H = \gamma H_i$, $\omega_M = \gamma 4\pi M_\mathrm{s}$, $\lambda_{\textrm{ex}}$ is the exchange length, and $\sin^2\theta_q = 1 - (q_{_\parallel}\cos\Theta_M + \kappa\sin\Theta_M)^2/q^2$, where $\theta_q$ is the angle between $\textbf{\textit{q}}$ and $\textbf{\textit{M}}$.
Equation\,(\ref{dispersion}) determines the allowed values of the perpendicular wavevector component $\kappa$, and is, effectively, a 6-th order equation in $\kappa$. In the frequency range of propagating magnon modes, two solutions of Eq.\,(\ref{dispersion}), $\kappa_{\mathrm{1}}$ and $\kappa_{\mathrm{2}}$, are real, while the other four solutions have non-zero imaginary parts.
The complex solutions of Eq.\,(\ref{dispersion}) have the absolute values $|\kappa| \sim 1/\lambda_{\mathrm{ex}}$, and describe surface corrections to the profile along the film normal of the considered spin-wave mode. An account of these evanescent partial waves is necessary for the analysis of boundary conditions at the film surfaces. However, their influence is limited only to a rather thin boundary layer of the width $\Delta\xi \approx 1/|\kappa| \sim\lambda_{\mathrm{ex}}$. In a magnetic film of a finite thickness $L$, the profile of a spin-wave mode is accurately described outside from a thin boundary layers of the width $\Delta\xi \approx 1/|\kappa| \sim\lambda_{\mathrm{ex}}$ by the sum of two frequency-degenerate partial waves with the real-valued perpendicular wavevector components $\kappa_{\mathrm{1}}$ and $\kappa_{\mathrm{2}}$. \looseness=-1

\smallskip
\noindent\textbf{Spin-wave spectra and mode profiles calculation.}
We consider a ferromagnetic film of finite thickness $L$ in the $\xi$ direction (see inset in Fig.\,\ref{FigExperiment}a), that is infinite in the two in-plane directions. We assume that a non-uniform spin-wave mode of vector amplitude $\textbf{m}(\xi)$ is propagating along the $\zeta$ axis lying in the film plane, and that this axis is parallel to the projection of the obliquely applied bias magnetic field $\textbf{\textit{H}}_0$ onto the film plane (see inset in Fig.\,\ref{FigExperiment}a):
\begin{equation} \label{eq2}
\mathbf{m}(\xi,\zeta,t) = \mathbf{m}(\xi) \exp [i ( \omega t-q_{_\parallel} \zeta )] \, ,
\end{equation}
where $\omega$ is the frequency of the spin-wave mode.	

Following the method used in  Ref.\,\cite{Kalinikos1986}, we expand the spin-wave mode profile along the film normal $\mathbf{m}(\xi)$ into an infinite series of complete orthogonal vector functions, that are the eigenfunctions of the second-order exchange differential operator and the exchange boundary conditions corresponding to zero surface anisotropy (unpinned surface spins):
\begin{equation} \label{eq3}
\mathbf{m}(\xi) \propto \sum_{n}  \mathbf{m}_n \cos [\frac{\pi n}{L} (\xi + \frac{L}{2})] \, .
\end{equation}

Using Eq.\,(\ref{eq3}) in the framework of the formalism of Ref.\,\cite{Kalinikos1986}, it is possible to reduce the Landau-Lifshitz equation describing the spin-wave dynamics in a magnetic film to the following infinite system of algebraic equations for the vector amplitudes $\mathbf{m}_n$ of the magnon modes:

\begin{equation} \label{eq4}
i \frac{\omega}{\omega_{\mathrm{M}}} \mathbf{m}_n = \sum_{n'} \hat{W}_{nn'} \mathbf{m}_{n'} \, ,
\end{equation}
where the square matrix $\hat{W}$ can be found by simple algebraic transformation of Eq.\,(22) from Ref.\,\cite{Kalinikos1986}.

Next, the dispersion characteristics $\omega(q_{_\parallel})$ and the profiles along the film normal $\mathbf{m}(\xi)$ of the spin-wave modes can be calculated as the eigenvalues and eigenvectors of the matrix $\hat{W}$ (see Eq.\,(\ref{eq4})).

\end{document}